\documentclass[twocolumn,prb,aps,superscriptaddress,showpacs]{revtex4}
\usepackage{graphicx}
\begin{document}
\title {Temperature dependence of the conductivity of the electronic
crystal}
\author{Shi-Jie Yang}
\affiliation{Department of Physics, Beijing Normal University,
Beijing 100875, China}
\author{Yue Yu}
\affiliation{Institute of Theoretical Physics, Chinese Academy of
Sciences, P.O. Box 2735, Beijing 100080, China}

\begin{abstract}
We study the temperature dependence of the conductivity of the 2D
electronic solid. In realistic samples, a domain structure forms
in the solid and each domain randomly orients in the absence of
the in-plane field. At higher temperature, the electron transport
is governed by thermal activation form of $\sigma_{xx}(T)\propto
e^{-\Delta_0/k_BT}$. The impurities will localize the electron
states along the edges of the crystal domains. At sufficient low
temperature, another transport mechanism called Mott's variable
range hopping mechanism, similar to that in a disorder insulator
takes effect. We show that as the temperature decreases, a
crossover from the fixed range hopping of the transport to the
variable range hopping of transport in the 2D electron system may
be experimentally observed.

\end{abstract}
 \maketitle

It was initially predicted by Wigner that two-dimensional (2D)
electrons crystallize into a triangular lattice in the low density
limit where the electron-electron interactions dominate over the
kinetic energy. In an ideally clean 2D system, the critical $r_s$
was presented to be $37\pm 5$ from quantum Monte Carlo simulations
\cite{Tanatar}. A strong magnetic field perpendicular to the 2D
plane can effectively localize electron wave functions while
keeping the kinetic energy controlled\cite{Yosh}. Since this
lessens the otherwise severe low-density condition, it is believed
that the Wigner crystal (WC) can be stabilized in a sufficiently
strong magnetic field\cite{Lau,Lam,Kuk,Santos}. Approximate
calculations\cite{CM1} have shown that the WC becomes the lowest
energy state when the filling factor $\nu < 1/6$ for $GaAs/AlGaAs$
electron system and around $\nu=1/3$ for the hole system.

As well-known, while the transport behavior of the WC is
characterized by non-linear I-V curves, the temperature dependence
of the conductivity of a WC is believed to be normal, namely, it
has an ordinary thermal activation form\cite{Chui,Li}. Moreover,
since the impurities pin the electronic crystal, a domain
structure forms in realistic samples\cite{Sher,Fil}. While the
electrons in a domain have an order as they are in the ideal
crystal, the orientations of the domains are random. Another role
played by the impurities is they localize the electron states
along the edges of the crystal domains. The electrons may hop
between the edges of the domains. In this case, the transport
behavior of the system is similar to that in a disordered
insulator. In this work, we will show that the transport of the
electrons in the 2D electronic solid obeys a generalized Mott's
variable range hopping theory\cite{mott} for a low temperature
$T\ll T_0$, while it obeys the ordinary thermal activation law for
$T\geq T_0$. We find that $T_0$ is in the experimentally reachable
regime if the sample parameters are properly chosen. This implies
that one may experimentally observe a different temperature
dependence of the conductivity at different temperature regimes.
As $T$ varies, the conductivity, according to our calculation, may
have a crossover from $\sigma_{xx}(T)\propto e^{-A/T^{1/2}}$ for
$T\ll T_0$ to a thermal activation form $\sigma_{xx}(T)\propto
e^{-\Delta_0/k_BT}$.

Pinning of the WC by impurities as a result of breaking of the
translational invariance has been widely
investigated\cite{Kuk,Millis}. In realistic samples a domain
structure is formed due to a finite impurity density. The
electrons in each domain are ordered as they are in the crystal.
Sherman\cite{Sher} and Fil\cite{Fil} had studied the angular
pinning and the domain structure of the electronic crystal
mediated by acoustic-phonon in $III-V$ semiconductor. In the
absence of the in-plane field, each domain orients randomly, just
like the domains in ferromagnets. It can be shown that an in-plane
field favors the domains to orient to the same
direction\cite{yang}. Hence the in-plane magnetic field may serve
as a tunable means to probe the orientation of the crystal. An
ideal electronic crystal is an insulator and the conductivity
$\sigma_{xx}\propto e^{-\Delta_0/k_BT}$. This thermal activation
form of the conductivity implies that the electrons are hopping
with a fixed range mechanism. It has been confirmed by experiments
with $\Delta_0$ typically of the order 1K \cite{jiang}. In a
realistic domain structure, however, the localized electrons may
hop between the edges of the randomly oriented domains. Since the
experimentally reachable temperature is as low as 10mK, the
variable range hopping mechanism \cite{mott} may work in this
temperature regime. In the following, we will calculate the
electron conductivity according to the different electron hopping
mechanisms and determine the characteristic temperature of the
crossover region $T_0$.

In the usual Anderson localization the envelope of the wave
function falls off exponentially as $\phi\sim e^{-r/\xi}$, where
$\xi$ is the localization length, while in the magnetic field the
electronic wave function of a perfect system is essentially a
Gaussian as $\phi\sim e^{-r^2/2l_B^2}$, where $l_B$ is the
magnetic length. In a slightly disordered system one can think
that some of the states will be pinned at certain isolated
impurity site. The mixing of these states due to
quantum-mechanical tunnelling leads to a simple exponential tail
in the wave function\cite{Li}. In a strong magnetic field, the
electrons condense into a crystal at lower filling factors. The
electrons are Coulomb localized. When the temperature is high
enough the transport is of the thermal activation form, which
implies that the electrons are hopping with a fixed range
mechanism\cite{Chui,Li}. The hopping range is determined by
$R_0=\sqrt{1/\pi n_I}$, where $n_I$ is the impurity density.
However, localized states by impurities may exist along the edges
of the domains of electronic crystal. When the temperature is
sufficiently low such that there is nearly no phonon with energy
to assist the electron making the nearest hopping, Mott's variable
hopping mechanism\cite{mott} allows the electrons hop a larger
distance $R>R_0$ to a state which has a smaller energy difference
$\Delta(R)$. In turn, the hopping conduction is determined by the
typical decay rate of the tails of the wave function. The hopping
probability is then given by
\begin{equation}
p\propto \exp[-R/\xi-\Delta/k_BT], \label{prb}
\end{equation}
where $R=|\vec R_i-\vec R_j|$ and $\Delta$ is the activation
energy.

For non-interacting electrons, Mott hopping with an approximately
constant density of states at the Fermi energy gives,
\begin{equation}
\rho_{xx}(T)=\rho_0(T)exp(r_c/\xi)=\rho_0(T)exp(A_0/T)^{1/3},
\end{equation}
where $r_c$ is a characteristic hopping length, which in this
regime is equal to the Mott hopping length. However, this simple
treatment does not work for the WC because of the strong Coulomb
interaction between electrons\cite{Pollak}. The Coulomb gap
depresses the density of states near the Fermi surface
\cite{efros1,efros2,YM,Chan}. Efros {\it et al} \cite{efros} had
derived the density of states near the Fermi surface $N(E)\propto
|\Delta E|= |E-E_F|$. The condition to find one state within a
circle of radius $R$ is given by
\begin{equation}
\pi R^2N(E_F)\Delta(R)=1. \label{cond}
\end{equation}
Substitute $N(E_F)$ by the available states near the Fermi
surface, {\it i.e.}, $N(E_F)\to \bar N(E_F)=\frac{1}{\Delta
E}\int_0^{\Delta E} dE N(E)$, and note that $|\Delta E|\propto
1/R$, we get
\begin{equation}
\Delta(R)\sim \frac{\hbar v_F}{ R},
\end{equation}
where $v_F$ is the Fermi velocity. Put this $R$-dependent energy
difference into formula (\ref{prb}) and maximizing $p$, one finds
the optimal hopping range $R=\bar R$ and the maximum of the
probability are given by
\begin{eqnarray}
\bar R^2= \frac{\hbar v_F \xi}{k_BT},~~p\propto
e^{-2\bar{R}/\xi}=e^{-A/T^{1/2}},
\end{eqnarray}
with $A=[\frac{4\hbar v_F}{k_B \xi}]^{1/2}$. The conductivity in
the variable range hopping is then\cite{efros1,efros2}
\begin{equation}
\sigma_{xx}\propto p\propto e^{-A/T^{1/2}}.
\end{equation}

The characteristic temperature $T_0$ above which the fixed range
hopping dominates is determined by $\bar R=R_0$, namely
\begin{eqnarray}
k_BT_0=\pi n_I\hbar v_F\xi=\pi n_I\cdot \frac{\hbar^2\xi}{m_bl_B}.
\label{t0}
\end{eqnarray}
In a strong magnetic field the decay length is comparable to the
cyclotron radius $\xi\sim R_c$\cite{Ebert,Fogler}. We find for
sample with $n_I\sim 1.0\times 10^8$cm$^{-2}$, $T_0\sim 40$mK .
This temperature is experimentally reachable. We anticipate the
different dependence of the conductivity in different temperature
regimes can be observed in future experiments.

Now, we briefly discuss the effect of the tilted field. Consider
an electron moving on a $x$-$y$ plane under the influence of a
strong magnetic field which is tilted an angle $\theta$ to the
normal, with $\vec{B}=(B\tan\theta,0,B)$. The electron is confined
in a harmonic potential $V(z)={1\over 2}m_b \Omega^2 z^2$ in the
$z$-direction, where $m_b$ is the band mass of the electron and
$\Omega$ the characteristic frequency. Such a quantum well has
been chosen to deal with many quantum Hall
systems\cite{Phillips,Jungwirth,yu} to substitute the realistic
potential which is either triangular \cite{cha} or
square\cite{mac}. We work in the "Landau gauge" by choosing the
vector potential $\vec{A}=\{0, x B_z-z B_x, 0\}$. The single
particle wave function for the lowest LL are:
\begin{eqnarray}
\phi_{X}(\vec{r})&=&{1\over \sqrt{L_y}}e^{-\frac{iXy}{l_B^2}}
    \Phi_0^{\omega_+}((x-X)\sin\tilde{\theta}+z\cos\tilde{\theta}) \nonumber\\
   & & \times \Phi_0^{\omega_-}((x-X)\cos\tilde{\theta}-z\sin\tilde{\theta}),
\label{one}
\end{eqnarray}
where $l_B$ is the magnetic length and $X$ is an integer multiple
of $2\pi l_B^2/L$. $\Phi_0^{\omega_\pm}$ is the harmonic
oscillator wave function in the lowest energy level corresponding
to the frequencies $\omega_\pm$ and
$\tan\tilde{\theta}=\frac{\omega_c^2}{\omega_+^2-\omega_c^2}\tan\theta$,
with the cyclotron frequency $\omega_c=eB/m_bc$. The frequencies
$\omega_\pm$ are given by\cite{yang2}
\begin{equation}
\omega_\pm^2={1\over 2}(\Omega^2+\frac{\omega_c^2}{\cos^2 \theta})
\pm\sqrt{{1\over 4}(\Omega^2-\frac{\omega_c^2}{\cos^2 \theta})^2
+\Omega^2 \omega_c^2 \tan^2\theta}.
\end{equation}

Obviously, the existence of an in-plane field deforms the electron
wave function. However, this wave function deformation does not
qualitatively change the electron hopping mechanism at a given
temperature. The major effect of the tilted field would be on the
variation of $T_0$. Fig. 1 illustrates the cohesive energy for two
typical configurations of the crystal orientation with respective
to the in-plane field: the $[100]$ or the $[110]$ direction
parallel to the in-plane field. The energy is always lower for the
case of the $[110]$ direction parallel to the in-plane field. The
applied in-plane field lowers the cohesive energy of the
electronic crystal and forces the domains align to the same
direction. Thus, the role of the in-plane field is to integrate
the domains into larger ones. In this way, the in-plane field
causes some of impurities to be irrelevant and therefore reduces
the effective impurity density. In determining $T_0$ from
(\ref{t0}), only the relevant impurities should be counted in.
Hence, one can replace $n_I$ by an effective impurity density
$n_I(B_\parallel)$. From eq. (\ref{t0}), we see that $T_0$ is
sensitive to $n_I(B_\parallel)$. Therefore, it is possible to
observe the influence of the in-plane magnetic field on the
characteristic temperature $T_0$ under proper parameters as the
tilting angle varies.

In conclusion, the temperature dependence of conductivity is
explored for the Wigner crystal in 2-dimensional electrons under a
strong magnetic field. We argued that there are domains of
electronic crystal in a realistic sample and predicted that the
temperature dependence of the transport behavior may be different
in different temperature regimes. We found that the conductivity
experiences a crossover from the fixed range hopping to the
variable range hopping mechanism as the temperature varied. The
crossover temperature $T_0$ is reachable under present
experimental technique. Finally, the possible effect on the
crossover region by the in-plane magnetic field is discussed.

This work was supported in part by the NSF of China.

\centerline {Figure Captions} Figure 1 The cohesive energy of the
WC versus the tilting angle for two configurations of the crystal
orientation. The real line: $[110]$ parallel to the in-plane field
(as shown in the inset); The broken line: $[100]$ parallel to the
in-plane field.

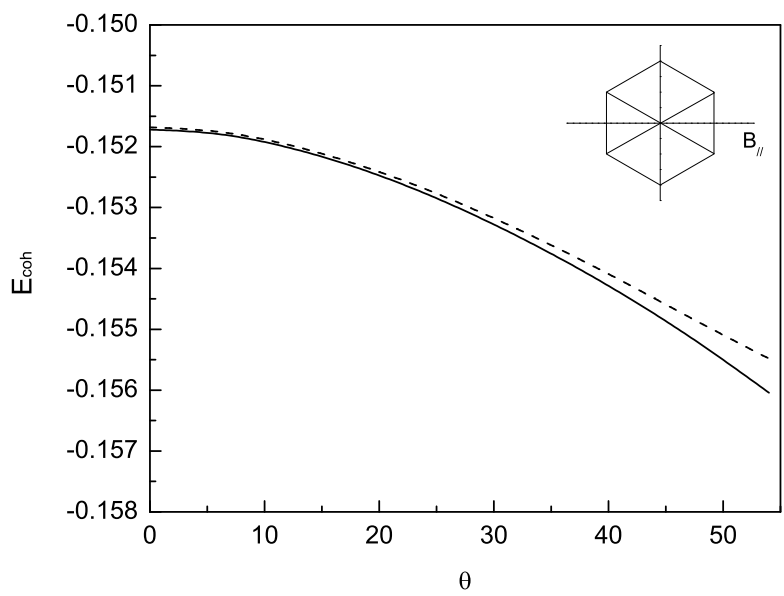
\end{document}